\documentclass[aps,prl,twocolumn,showpacs,superscriptaddress]{revtex4}
\usepackage{amsmath}

\usepackage{array}
\usepackage{graphicx}
\usepackage{dcolumn}
\usepackage{amsmath}
\usepackage{enumerate,amsthm,amssymb,color}
\usepackage{bbm}
\usepackage{bm}
\usepackage[french, english]{babel}
\usepackage{synttree}
\usepackage{multirow}

\newcommand{\tr}{\mathrm{ tr }}

\newcommand{\ket}[1]{|#1\rangle}
\newcommand{\bra}[1]{\langle#1|}
\newcommand{\ketbra}[2]{|#1\rangle\langle#2|}

\newcommand{\COMMENT}[1]{}

\begin{document}

\title{Nonlocality and conflicting interest games}

\author{Anna Pappa}  \affiliation{LTCI, CNRS -- T\'el\'ecom ParisTech, Paris, France}\affiliation{LIAFA, CNRS -- Universit\'e Paris 7, Paris, France}
\author{Niraj Kumar} \affiliation{Indian Institute of Technology, Kanpur, India}
\author{Thomas Lawson} \affiliation{LTCI, CNRS -- T\'el\'ecom ParisTech, Paris, France}
\author{Miklos Santha} \affiliation{LIAFA, CNRS -- Universit\'e Paris 7, Paris, France}\affiliation{CQT, National University of Singapore, Singapore}
\author{Shengyu Zhang} \affiliation{Dept of Computer Science and Engineering and ITCSC, The Chinese University of Hong Kong, Hong Kong}
\author{Eleni Diamanti}  \affiliation{LTCI, CNRS -- T\'el\'ecom ParisTech, Paris, France}
\author{Iordanis Kerenidis}  \affiliation{LIAFA, CNRS -- Universit\'e Paris 7, Paris, France}\affiliation{CQT, National University of Singapore, Singapore}
\date{\today}

\begin{abstract}
Nonlocality enables two parties to win specific games with probabilities strictly higher than allowed by any classical theory. Nevertheless, all known such examples consider games where the two parties have a common interest, since they jointly win or lose the game. The main question we ask here is whether the nonlocal feature of quantum mechanics can offer an advantage in a scenario where the two parties have conflicting interests. We answer this in the affirmative by presenting a simple conflicting interest game, where quantum strategies outperform classical ones. Moreover, we show that our game has a fair quantum equilibrium with higher payoffs for both players than in any fair classical equilibrium. Finally, we play the game using a commercial entangled photon source and demonstrate experimentally the quantum advantage.
\end{abstract}


\maketitle

Nonlocality is one of the most important and elusive properties of quantum mechanics, where two spatially separated observers sharing a pair of entangled quantum bits can create correlations that cannot be explained by any local realistic theory. More precisely, Bell \cite{Bell:64} showed that there exist scenarios where correlations between any local hidden variables can be shown to satisfy specific constraints (known as Bell inequalities), while these constraints can nevertheless be violated by correlations created by quantum systems. 

An equivalent way of describing Bell test scenarios is in the language of nonlocal games.
The best-known example is the CHSH game \cite{CHSH:prl69}: Alice and Bob, who are spatially separated and cannot communicate, receive an input bit $x$ and $y$ respectively and must output bits $a$ and $b$ respectively, such that the outputs are different if both input bits are equal to 1, and the same otherwise. It is well known that the probability over uniform inputs that they jointly win this game when they a priori share classical resources is $0.75$, while if they share and appropriately measure a pair of maximally entangled qubits, they can jointly win the game with probability $\cos^2 \pi/8 > 0.75$. The classical value $0.75$ corresponds to the upper bound of a Bell inequality and the CHSH game provides an example of a Bell inequality violation, since there exist quantum strategies that violate this bound.

Looking at Bell inequalities through the lens of games has been very useful in practice, including in cryptography \cite{ABG:prl07,BRG:arxiv13} and quantum information \cite{RUV:nat13}, where, for example, quantum mechanics offers stronger than classical security guarantees in quantum key distribution or verification protocols.
Recently, Brunner and Linden made the connection between Bell test scenarios and games with incomplete information more explicit and provided examples of such games where quantum mechanics offers an advantage \cite{brunlin:nc13}. A game with incomplete information (or Bayesian game) is a game where the two parties receive some input unknown to the other party \cite{harsanyi67}. We remark that without more restrictions, quantum mechanics only offers advantages for incomplete information games, \emph{i.e.}, when the parties receive inputs or, in other words, when there are more than a single measurement setting for each party \cite{zhang:11}.

There are two general types of games depending on whether the interests of the players are common or  conflicting \cite{osborn03}. A typical example of common interest games is when the drivers of vehicles decide on which side of the road they will drive.
When they both decide to drive on the same side, they win 1 point, while when they decide on different sides, they lose 1 point (because they crash their cars). In this game, it is easy to see that the outcomes ``both drive on the right'' and ``both drive on the left'' are equilibria, meaning that no party can increase their payoff by deviating unilaterally. Moreover, both players equally prefer each of the equilibria, hence there is no conflict on which one to choose. Another example is the CHSH game, where we can assume that both Alice and Bob win 1 point if the parity of their outputs is equal to the logical AND of their inputs, and they lose 1 point otherwise (see Example 1 in \cite{brunlin:nc13}). In fact, other known nonlocal games, including the GHZ-Mermin game \cite{mermin:ajp90}, the Magic Square Game \cite{Mermin:prl65.3373,Peres:pl90}, the Hidden Matching game \cite{BJK:stoc04,BRS:ccc11}, Brunner and Linden's three games \cite{brunlin:nc13}, are all examples of common interest games \footnote{Note that in all games where Alice and Bob's utility functions are the same, there can be no conflicts. The only example of game where the utility function is different is Example 2 in \cite{brunlin:nc13}; nevertheless, if we calculate the equilibria of this game then we can see that there are only two equilibria, in both of which the average payoff of Alice is 0 and that of Bob is 2. Hence this is not a conflicting interest game.}.

There is a second important type of games, called conflicting interest games, where the interests of the players differ. A typical example is the Battle of the Sexes, where Alice and Bob want to meet on Saturday night, but Alice prefers the ballet, while Bob prefers the theater.
For example, when they both go to the ballet, Alice wins 2 points and Bob 1, when they go to the theater, Alice wins 1 point and Bob 2, and when they go to different places, they do not get any points. In this game, the outcomes ``both go to the ballet'' and ``both go to the theater'' are equilibria, meaning that no party can increase their payoff by deviating unilaterally. However, each party prefers a different equilibrium; hence, how can they choose one of them and resolve the conflict? One way is to provide a common advice from a trusted referee. For example, both parties receive a uniformly random coin and go to the ballet when the coin is heads and to the theater when the coin is tails. This classical strategy with advice is, in fact, a fair, correlated equilibrium \cite{aumann:jme74}.

There have been numerous examples where quantum mechanics offers an advantage for common interest games, where the players either jointly win or jointly lose. For example, in the scenario of device independent key distribution or randomness extraction, the two boxes that participate in the Bell test are provided by a common adversary and they need to coordinate in order to jointly violate a Bell inequality. In the case of conflicting interest games, however, it is not a priori clear if such an advantage can be offered. For example, fundamental cryptographic games with competing players, such as coin flipping or bit commitment, remain impossible even in the presence of quantum resources \cite{lochau98,mayers}.

The main question we address in this Letter is whether the nonlocal feature of quantum mechanics can indeed offer an advantage in a scenario where the two non-communicating parties that participate in a Bell test scenario have conflicting interests. We can also restate our question in the language of games: are there conflicting interest games where quantum advice offers an advantage compared to classical advice? We answer in the affirmative by presenting a simple incomplete information game with conflicting interests, where there exist quantum strategies with average payoff for the two parties strictly higher than that allowed by any classical strategy. Moreover, we show that there exist fair, quantum equilibria with strictly higher payoffs than in any classical fair, correlated equilibrium. This is the first example, to our knowledge, where the nonlocal feature of quantum mechanics has been used to resolve a conflict between two parties in a way that is advantageous for both parties simultaneously. Finally, the simplicity of our game enables us to demonstrate experimentally these quantum strategies using a commercial entangled photon source and confirm that the average payoff of the players is strictly higher than classically possible.

\vspace{0.2cm}

\noindent
\emph{The conflicting interest Bayesian game}. We define a Bayesian game in the two-party framework (for a more general definition refer to \cite{osborn03}). It comprises of:
\begin{itemize}
\item Two players, Alice (A) and Bob (B).
\item A set $\mathcal{X}=\mathcal{X}_A\otimes \mathcal{X}_B$ of pairs of types/inputs $x=(x_A,x_B)$, where $x_A\in\mathcal{X}_A$, $x_B\in\mathcal{X}_B$, which follow a probability distribution $P:\mathcal{X}\rightarrow [0,1]$.
\item A set $\mathcal{Y}=\mathcal{Y}_A\otimes \mathcal{Y}_B$ of pairs of actions/outputs $y=(y_A,y_B)$, where $y_A\in\mathcal{Y}_A$, $y_B\in\mathcal{Y}_B$.
\item A utility function $u_i:\mathcal{X}_A\times \mathcal{X}_B\times \mathcal{Y}_A \times \mathcal{Y}_B\rightarrow \mathbb{R}$, which determines the gain for each player depending on the types and actions of both players.
\end{itemize}

\begin{figure}[tb]
\centering
\includegraphics[scale=0.25]{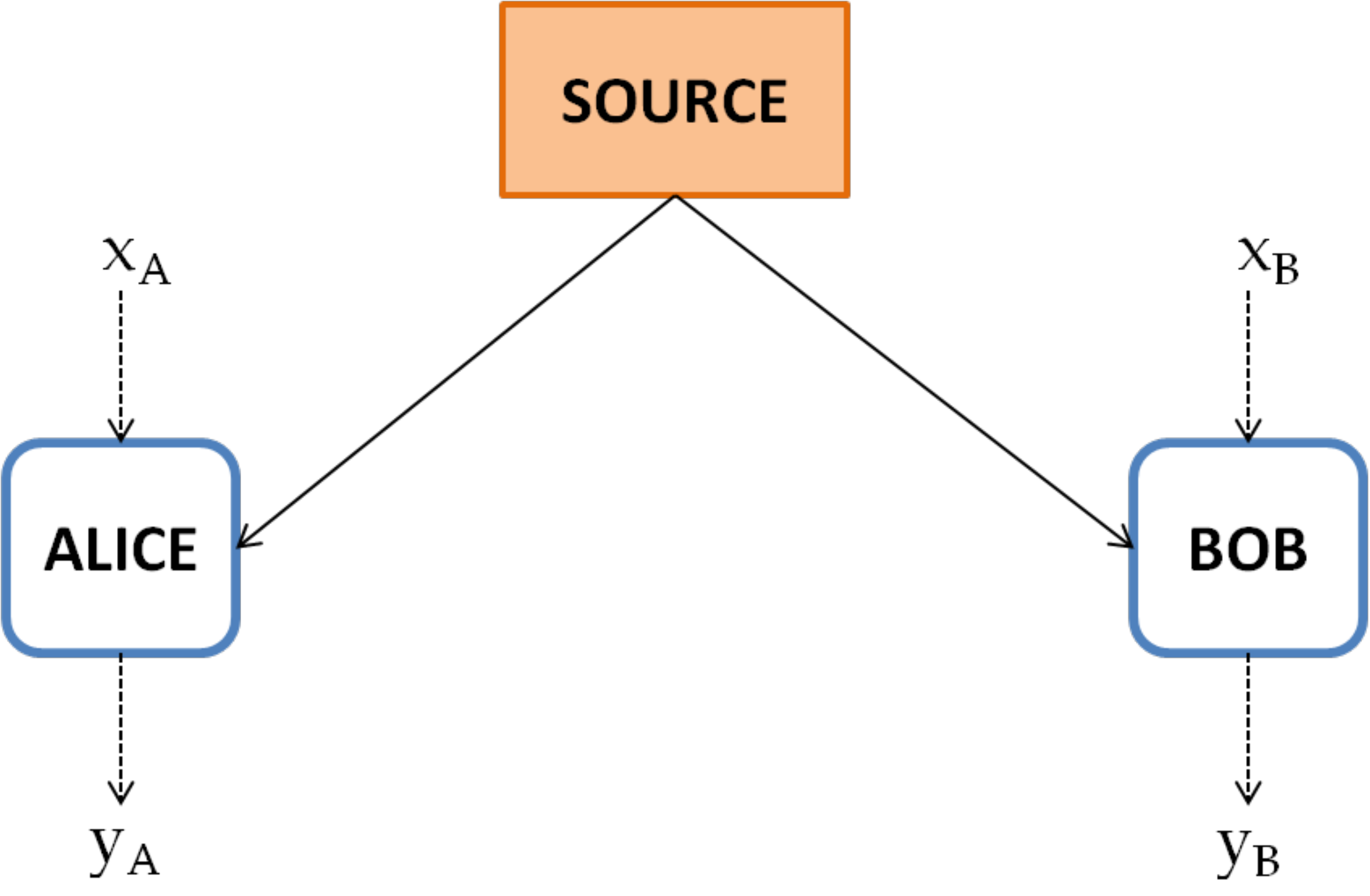}
\caption{Bayesian game configuration for two players.}\label{Bay_Game}
\end{figure}

In general, the game is played as shown in Fig. \ref{Bay_Game}. Each player $i\in\{A,B\}$ acquires a type $x_i$ according to the probability distribution $P$. We also consider the case where they receive an advice from a source that is independent of the inputs $x_i$, and that can be classical or quantum. Finally, they decide on their action/output $y_i$, according to a chosen strategy. Each player $i\in\{A,B\}$ is interested in maximizing his or her average payoff $F_i$:
\begin{equation}\label{avgpayoff}
F_i=\sum_{(x,y)\in\mathcal{X} \times \mathcal{Y} }   P(x)\Pr{(y|x)}u_i(x,y),\\
\end{equation}
where $\Pr{(y|x)}$ is the probability the players choose actions $y=(y_A,y_B)$ given their types were $x=(x_A,x_B)$ and depends on the advice and the chosen strategies.

In the case of classical advice, we define a correlated strategy $c_i:\mathcal{X}_i\otimes\Omega_i\rightarrow\mathcal{Y}_i$, where $\Omega_i$ is the space of advice given to player $i$ by the source and the source chooses the advice $r=(r_A,r_B)$ from the space $\Omega=\Omega_A\times\Omega_B$ following a probability distribution $Q$. Given a type $x_i$ and an advice $r_i$, player $i$ performs the action $y_i=c_i(x_i,r_i)$. If both players follow a correlated strategy $c=(c_A,c_B)$, the average payoff for player $i\in\{A,B\}$ becomes:
\begin{equation*}
F_i(c)=\sum_{\substack{x\in\mathcal{X}\\r\in\Omega}}  P(x) Q(r) u_i(x_A,x_B,c_A(x_A,r_A),c_B(x_B,r_B))
\end{equation*}
A correlated strategy $c$ is a correlated equilibrium \cite{aumann:jme74} if player $i$ cannot gain a higher payoff by changing his or her strategy unilaterally.
\COMMENT{
\[
F_i(c_i,c_{\neg{i}})\geq F_i(c'_i,c_{\neg{i}}),~~~~\forall i\in\{A,B\}
\]
}

In the case of quantum advice, we define a quantum strategy $\mathcal{M}=(\mathcal{A},\mathcal{B},\ket{\Psi})$, with $\mathcal{A}=(\mathcal{A}_0,\mathcal{A}_1)$ and $\mathcal{B}=(\mathcal{B}_0,\mathcal{B}_1)$, that consists of the players applying respectively the observables $\mathcal{A}_{x_A}=\{A_{x_A}^0,A_{x_A}^1\}$ and $\mathcal{B}_{x_B}=\{B_{x_B}^0,B_{x_B}^1\}$ on the shared quantum state/advice $\ket{\Psi}$. The probability of the two players outputting $y$ given $x$, is $\Pr{(y|x)}=\bra{\Psi}M_x^y\ket{\Psi}$, where $M_x^y=A_{x_A}^{y_A}\otimes B_{x_B}^{y_B}$. The average payoff for player $i\in\{A,B\}$ becomes:
\begin{equation*}\label{quantumpayoff}
F_i(\mathcal{M})=\sum_{ x\in\mathcal{X}, y\in\mathcal{Y} } P{(x)}\bra{\Psi}M_x^y\ket{\Psi} u_i(x,y)\\
\end{equation*}
A quantum strategy $\mathcal{M}$ is a quantum equilibrium if no player can gain a higher payoff by choosing a different strategy unilaterally.

For each combination of the players' types, the utility functions $u_A$ and $u_B$  can be viewed as a table: the rows correspond to the actions $y_A$ and the columns to the actions $y_B$. The numbers in each cell are the players' utilities ($u_A, u_B$) depending on their types and actions. In case the utilities differ for different types of the two players, then we need to introduce more than one table.

Our game is a combination of the Battle of the Sexes and the CHSH game. The utilities are defined in Table \ref{table} depending on the logical AND of the types of the players. We have normalized the utilities of the game to be in $[0,1]$. The dependence of the utilities on the types is similar to the CHSH game, where the players need to coordinate, except for the case where their types are both 1, in which case they need to anti-coordinate. However, similar to the Battle of the Sexes, the interests of the players are conflicting, since whenever at least one type is 0, the first player prefers the action $(0,0)$ and the second player prefers the action $(1,1)$. This is in stark constrast to the usual CHSH game, where the players have a common interest, since their utilities are always the same. In the following we consider that the types of the players are chosen uniformly at random.

\COMMENT{
\begin{table}[h!]
        \centering
        \renewcommand{\arraystretch}{1.4}
        \begin{tabular}{| c |c | c |}
     \hline
            & \makebox[6em]{$y_B=0$} & \makebox[6em]{$y_B=1$} \\\hline
     \makebox[6em]{$y_A=0$} & (1,1/2) & (0,0) \\\hline
    $y_A=1$ & (0,0) & (1/2,1) \\
    \hline
    \end{tabular}
    \caption{$x_A\land x_B=0$}\label{table1}\vspace{3mm}
    \begin{tabular}{| c |c | c |}
    \hline
           & \makebox[6em]{$y_B=0$} & \makebox[6em]{$y_B=1$} \\\hline
   \makebox[6em]{$y_A=0$} & (0,0) & (3/4,3/4) \\\hline
   $y_A=1$  &  (3/4,3/4) & (0,0) \\
    \hline
    \end{tabular}
\caption{$x_A\land x_B=1$}\label{table2}
\end{table}
}

\vspace{0.2cm}

\begin{table}[h!]
        \centering
        \renewcommand{\arraystretch}{1.4}
        \begin{tabular}{| c |c | c | c | c |}
     \cline{2-5}
   \multicolumn{1}{c}{}  & \multicolumn{2}{|c|}{ \makebox[4.7em]{$x_A\land x_B=0$}}& \multicolumn{2}{c|}{ \makebox[4.7em]{$x_A\land x_B=1$}} \\\hline
            & \makebox[4.7em]{$y_B=0$} & \makebox[4.7em]{$y_B=1$}& \makebox[4.7em]{$y_B=0$} & \makebox[4.7em]{$y_B=1$} \\\hline
     \makebox[4.7em]{$y_A=0$} & (1,1/2) & (0,0) & (0,0) & (3/4,3/4)\\\hline
    $y_A=1$ & (0,0) & (1/2,1) &  (3/4,3/4) & (0,0)\\
    \hline
    \end{tabular}
    \caption{Payoff table depending on the players' types.}\label{table}
 \end{table}

\noindent
\emph{Classical strategies}. We start by examining the equilibria in the absence of advice. There are three of them, a fair one, where both Alice and Bob have average payoff $9/16$; one where Alice receives $11/16$ and Bob $7/16$; and a third one where Alice receives $7/16$ and Bob $11/16$. It is clear that this is a conflicting interest game, since Alice prefers the second equilibrium and Bob the third.
Let us now examine classical correlated strategies, where in the beginning the source gives to each player an advice in the form of a bit $r_i$ drawn from some distribution independent of the types. Note that since there are only two possible actions per player, advice longer than a single bit does not increase the players' payoffs.

In our setting of classical advice and finite number of possible strategies, the set of all possible pairs of payoffs $(F_A,F_B)$ forms a convex polytope in $\mathcal{R}^2$. We can therefore examine all possible strategies and see that the average payoff for any player cannot exceed the value $3/4$ and, moreover, the two players cannot have their average payoff functions achieve their maximum at the same time; when $F_A = 3/4$, it holds that $F_B = 3/8$ and equivalently the other way around. We can finally verify that:
\begin{equation}\label{inequality}
F_A+F_B\leq \frac{9}{8} = 1.125
\end{equation}
As we mentioned earlier, there exists a fair classical equilibrium that provides average payoffs $F_A = F_B = 9/16$: in this case, Alice outputs her type and Bob the complement of his. There also exist several other correlated equilibria, depending on the probability distribution $Q$ of the advice, which of course satisfy Inequality (\ref{inequality}).

\vspace{0.2cm}
\noindent
\emph{Quantum strategies}. We first consider the case where the two players share a maximally entangled state, $\ket{\phi^+}=\frac{1}{\sqrt{2}}(\ket{00}+\ket{11})$. Following the analysis of the CHSH game by Cleve et al \cite{Cleve:CC04}, if the players use the following projective measurements according to their inputs:
\begin{eqnarray}\label{eq:bases}
\mathcal{A}_0^a&=&\ketbra{\phi_a(0)}{\phi_a(0)},~~~~~\mathcal{A}_1^a=\ketbra{\phi_a(\frac{\pi}{4})}{\phi_a(\frac{\pi}{4})}\nonumber\\
\mathcal{B}_0^b&=&\ketbra{\phi_b(\frac{\pi}{8})}{\phi_b(\frac{\pi}{8})},~~~~\mathcal{B}_1^b=\ketbra{\phi_b(-\frac{\pi}{8})}{\phi_b(-\frac{\pi}{8})}~~~~~
\end{eqnarray}
where $a,b\in\{0,1\}$ and $\phi_0(\theta)=\cos{\theta}\ket{0}+\sin{\theta}\ket{1}$, $\phi_1(\theta)=-\sin{\theta}\ket{0}+\cos{\theta}\ket{1}$, then $\Pr{(y_A,y_B|x_A,x_B)}=\frac{1}{2}\tr{(\mathcal{A}_{x_A}^{y_A},\mathcal{B}_{x_B}^{y_B})}=\frac{1}{2}\cos^2{\frac{\pi}{8}}$. The average payoff of player $i\in\{A,B\}$ is:
\begin{equation*}\label{eq:Fiquantum}
F_i=\frac{1}{8}\cos^2{\frac{\pi}{8}}\sum_{(x,y)\in\mathcal{X}\times \mathcal{Y} }  u_i(x,y)=\frac{3}{4} \cos^2{\frac{\pi}{8}} = 0.64
\end{equation*}

\begin{figure}[tb]
\centering
\includegraphics[scale=0.6]{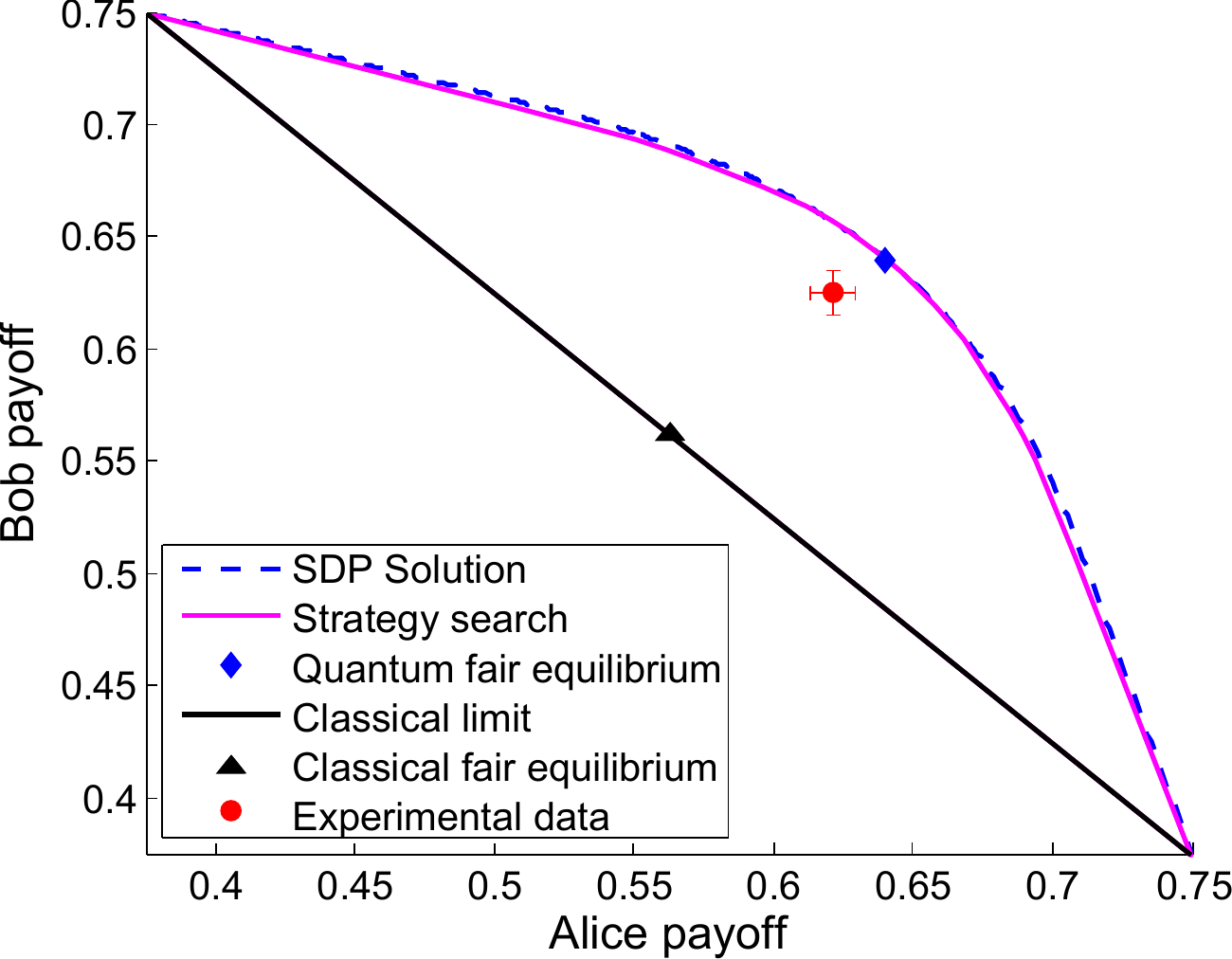}
\caption{Comparison of classical and quantum strategies for the Bayesian game with conflicting interests defined by Table \ref{table}. These strategies include the classical and quantum fair equilibria points. The experimentally obtained payoff is strictly higher than the classical bound.}\label{clasquantum}
\end{figure}

For each player, the ratio of the quantum over classical payoff for the quantum strategy $\mathcal{M}=(\mathcal{A},\mathcal{B},\ket{\phi^+})$, is the same as in the case of the CHSH game, in other words, our equilibrium corresponds to the Tsirelson bound for the maximum violation of the CHSH game.
We also prove that $\mathcal{M}$ is a quantum equilibrium: to this end, we use semidefinite programming (SDP) and verify that while keeping one player's strategy fixed to that defined by $\mathcal{M}$, the optimal strategy of the other player is indeed the one prescribed by $\mathcal{M}$. It is important to note that all Bell states lead to a quantum equilibrium, by appropriately rotating the measurement bases.
We also observe that the strategy $\mathcal{M}$ is fair, since it gives equal average payoffs. The joint payoff takes the value $F_A + F_B = 1.28$, which implies that, in our conflicting interest game, there exist fair quantum equilibria where the parties jointly have strictly higher payoff than in any classical fair, correlated equilibrium.  In addition, when we optimize over all quantum strategies that achieve fair average payoffs, by ranging over the joint state and measurement operators, we conclude that the above equilibrium is the optimal fair quantum equilibrium. Furthermore, we can find a whole range of quantum strategies (not equilibria) where the joint payoff of the players is strictly higher than classically possible. Finally, we show that these explicit strategies are very close to the optimal ones, by providing an upper bound that corresponds to the second level of the SDP hierarchy in  \cite{NPA:prl2007,NPA:njp2008,PNA:siam2010,KRT:siam2010}. In Fig. \ref{clasquantum} we have plotted the classical bound of Inequality (\ref{inequality}), quantum strategies that achieve higher joint payoff, the SDP upper bound for the optimal joint payoff for any quantum strategy, as well as the classical and quantum fair equilibria.

\vspace{0.2cm}
\noindent
\emph{Experimental demonstration}. The main component of the conflicting interest game that we have defined is the CHSH game. This simple setting allows us to demonstrate our game in practice using the commercial entangled photon source quED by QuTools \cite{QuTools}, which generates polarization entangled photon pairs in the state $\ket{\phi^+}=\frac{1}{\sqrt{2}}(\ket{\text{HH}}+\ket{\text{VV}})$ at a wavelength of 810 nm using type-I spontaneous parametric down-conversion. For each run of the game, we measure the polarization of the photons in the bases defined in (\ref{eq:bases}), where the measurement settings (types) are selected using a rotating quarter waveplate and a polarization filter placed at the path of each photon of the pair. Detection events are registered using the control and readout unit of quED, which includes a twin silicon avalanche photodiode module and a coincidence counter.

We would like to demonstrate a quantum strategy, such that the sum of the average payoffs of the two parties for a single run of the game is strictly higher than classically possible. For this, we take a large number of independent runs of the game in order to estimate each player's average payoff with high confidence. For each configuration, we record the single counts for each detector as well as coincidence counts. We also correct for the accidental coincidence counts.

We measure the fidelity of the generated state with respect to the maximally entangled state $\ket{\phi^+}$ to be equal to 0.925. The obtained violation of the CHSH inequality is 2.645, which corresponds to a probability of winning the usual CHSH game of 0.83. This is slightly lower than the maximum probability of success $\cos^2 \pi/8 \approx 0.85$, and is due to the imperfect fidelity of our source's state.

From the registered detection events for the measurement settings ($x_A,x_B$), we can compute the probabilities $\Pr{(y_A,y_B|x_A,x_B)}$, and hence the average payoff functions for the two players. We find:
\begin{equation*}\label{eq:data}
F_A = 0.621, \mbox{ }\mbox{ } F_B = 0.625.
\end{equation*}
The joint payoff is then $F_A + F_B = 1.246$, which is well above the classical bound of Inequality (\ref{inequality}) and slightly below the maximum value allowed by quantum strategies. Note that even if we do not correct for the effect of the accidental coincidences on the detection events, the obtained payoffs still largely surpass the classical bound.

The experimentally obtained payoff is plotted in Fig. \ref{clasquantum}, together with the classical bound and the space of the average payoffs for the quantum strategies. We also show the classical and quantum fair equilibria points. Our implementation demonstrates a payoff that is strictly higher than that obtained by any classical strategy, however, due to the non unit fidelity of our entangled state, this payoff does not reach the fair quantum equilibrium value.

Our results show that the implementation does not define a perfectly fair strategy since there is a small difference between the average payoffs of the two players, which is due to experimental imperfections. This form of bias can be eradicated by adding a shared uniformly random bit $r_0$ in the advice and having the players change their action if $r_0 = 1$ and proceed normally if $r_0=0$. In this case, the only bias that will be left in the experiment will come from the random number generator, which can be made vanishingly small.

\vspace{0.2cm}
\noindent
\emph{Discussion}. It is interesting to note that the game that we have proposed can also be seen as a Bell inequality, the one coming from Inequality (\ref{inequality}). The difference between almost all previous Bell inequalities that arise from games is that in our case the payoff functions of the players are not equal and hence, the Bell inequality arises only when we take the average of the two payoff functions and not just one of them (see also Example 2 in \cite{brunlin:nc13}). One can also ask what the payoffs are in case the players receive stronger, non-signaling correlations. It is straightforward to see that in this case, the fair equilibrium has average payoff $3/4$ for each player.

It would be interesting to find more conflicting interest games where quantum mechanics offers an advantage, for example when larger dimensions are used or in a multiparty setting \cite{benhay:pra01}. Finally, a more general question is whether every Bell inequality can be transformed into a conflicting interest game with the same maximal violation.

\vspace{0.2cm}
\noindent
\emph{Acknowledgments}. We thank Stefano Pironio for pointing out an improved upper bound on our game's payoffs that comes from the second level of the SDP hierarchy defined in \cite{NPA:prl2007}. We acknowledge financial support from the City of Paris through the project CiQWii. I.K. was supported by the ERC project QCC. A.P. and T.L. acknowledge support from Digiteo. S.Z. was partially supported by Hong Kong GRF Projects CUHK419011 and CUHK419413\\


\begin{thebibliography}{24}
\expandafter\ifx\csname natexlab\endcsname\relax\def\natexlab#1{#1}\fi
\expandafter\ifx\csname bibnamefont\endcsname\relax
  \def\bibnamefont#1{#1}\fi
\expandafter\ifx\csname bibfnamefont\endcsname\relax
  \def\bibfnamefont#1{#1}\fi
\expandafter\ifx\csname citenamefont\endcsname\relax
  \def\citenamefont#1{#1}\fi
\expandafter\ifx\csname url\endcsname\relax
  \def\url#1{\texttt{#1}}\fi
\expandafter\ifx\csname urlprefix\endcsname\relax\def\urlprefix{URL }\fi
\providecommand{\bibinfo}[2]{#2}
\providecommand{\eprint}[2][]{\url{#2}}

\bibitem[{\citenamefont{Bell}(1964)}]{Bell:64}
\bibinfo{author}{\bibfnamefont{J.~S.} \bibnamefont{Bell}},
  \bibinfo{journal}{Physics \textbf{1} (3): 195--200}
  (\bibinfo{year}{1964}).

\bibitem[{\citenamefont{Clauser et~al.}(1969)\citenamefont{Clauser, Horne,
  Shimony, and Holt}}]{CHSH:prl69}
\bibinfo{author}{\bibfnamefont{J.~F.} \bibnamefont{Clauser}},
  \bibinfo{author}{\bibfnamefont{M.~A.} \bibnamefont{Horne}},
  \bibinfo{author}{\bibfnamefont{A.}~\bibnamefont{Shimony}}, \bibnamefont{and}
  \bibinfo{author}{\bibfnamefont{R.~A.} \bibnamefont{Holt}},
  \bibinfo{journal}{Phys. Rev. Lett.} \textbf{\bibinfo{volume}{23}},
  \bibinfo{pages}{880} (\bibinfo{year}{1969}).

\bibitem[{\citenamefont{Acin et~al.}(2007)\citenamefont{Acin, Brunner, Gisin,
  Massar, Pironio, and Scarani}}]{ABG:prl07}
\bibinfo{author}{\bibfnamefont{A.}~\bibnamefont{Acin}},
  \bibinfo{author}{\bibfnamefont{N.}~\bibnamefont{Brunner}},
  \bibinfo{author}{\bibfnamefont{N.}~\bibnamefont{Gisin}},
  \bibinfo{author}{\bibfnamefont{S.}~\bibnamefont{Massar}},
  \bibinfo{author}{\bibfnamefont{S.}~\bibnamefont{Pironio}}, \bibnamefont{and}
  \bibinfo{author}{\bibfnamefont{V.}~\bibnamefont{Scarani}},
  \bibinfo{journal}{Phys. Rev. Lett.} \textbf{\bibinfo{volume}{98}},
  \bibinfo{pages}{230501} (\bibinfo{year}{2007}).

\bibitem[{\citenamefont{Brandao et~al.}(2013)\citenamefont{Brandao, Ramanathan,
  Grudka, Horodecki, Horodecki, and Horodecki}}]{BRG:arxiv13}
\bibinfo{author}{\bibfnamefont{F.}~\bibnamefont{Brandao}},
  \bibinfo{author}{\bibfnamefont{R.}~\bibnamefont{Ramanathan}},
  \bibinfo{author}{\bibfnamefont{A.}~\bibnamefont{Grudka}},
  \bibinfo{author}{\bibfnamefont{K.}~\bibnamefont{Horodecki}},
  \bibinfo{author}{\bibfnamefont{M.}~\bibnamefont{Horodecki}},
  \bibnamefont{and}
  \bibinfo{author}{\bibfnamefont{P.}~\bibnamefont{Horodecki}},
  \bibinfo{journal}{Preprint arXiv:1310.4544 [quant-ph]}
  (\bibinfo{year}{2013}).

\bibitem[{\citenamefont{Reichardt et~al.}(2013)\citenamefont{Reichardt, Unger,
  and Vazirani}}]{RUV:nat13}
\bibinfo{author}{\bibfnamefont{B.}~\bibnamefont{Reichardt}},
  \bibinfo{author}{\bibfnamefont{F.}~\bibnamefont{Unger}}, \bibnamefont{and}
  \bibinfo{author}{\bibfnamefont{U.}~\bibnamefont{Vazirani}},
  \bibinfo{journal}{Nature} \textbf{\bibinfo{volume}{496}},
  \bibinfo{pages}{456} (\bibinfo{year}{2013}).

\bibitem[{\citenamefont{Brunner and Linden}(2013)}]{brunlin:nc13}
\bibinfo{author}{\bibfnamefont{N.}~\bibnamefont{Brunner}} \bibnamefont{and}
  \bibinfo{author}{\bibfnamefont{N.}~\bibnamefont{Linden}},
  \bibinfo{journal}{Nat. Commun.} \textbf{\bibinfo{volume}{4}},
  \bibinfo{pages}{2057} (\bibinfo{year}{2013}).

\bibitem[{\citenamefont{Harsanyi}(1967/1968)}]{harsanyi67}
\bibinfo{author}{\bibfnamefont{J.~C.} \bibnamefont{Harsanyi}},
  \bibinfo{journal}{Management Science 14 (3): 159-183 (Part I), 14 (5):
  320-334 (Part II), 14 (7): 486-502 (Part III)}  (\bibinfo{year}{1967/1968}).

\bibitem[{\citenamefont{Zhang}(2012)}]{zhang:11}
\bibinfo{author}{\bibfnamefont{S.}~\bibnamefont{Zhang}}, in
  \bibinfo{journal}{Proceedings of the 3rd Innovations in Theoretical Computer Science (ITCS): 39--59}
  (\bibinfo{year}{2012}).

\bibitem[{\citenamefont{Osborne}(2003)}]{osborn03}
\bibinfo{author}{\bibfnamefont{M.~J.} \bibnamefont{Osborne}},
  \emph{\bibinfo{title}{An Introduction to Game Theory}}
  (\bibinfo{publisher}{Oxford University Press}, \bibinfo{year}{2003}).

\bibitem[{\citenamefont{Mermin}(1990{\natexlab{a}})}]{mermin:ajp90}
\bibinfo{author}{\bibfnamefont{N.~D.} \bibnamefont{Mermin}},
  \bibinfo{journal}{Am. J. Phys.} \textbf{\bibinfo{volume}{58}},
  \bibinfo{pages}{731} (\bibinfo{year}{1990}{\natexlab{a}}).

\bibitem[{\citenamefont{Mermin}(1990{\natexlab{b}})}]{Mermin:prl65.3373}
\bibinfo{author}{\bibfnamefont{N.~D.} \bibnamefont{Mermin}},
  \bibinfo{journal}{Phys. Rev. Lett.} \textbf{\bibinfo{volume}{65}},
  \bibinfo{pages}{3373} (\bibinfo{year}{1990}{\natexlab{b}}).

\bibitem[{\citenamefont{Peres}(1990)}]{Peres:pl90}
\bibinfo{author}{\bibfnamefont{A.}~\bibnamefont{Peres}},
  \bibinfo{journal}{Phys. Lett. A} \textbf{\bibinfo{volume}{151}},
  \bibinfo{pages}{107} (\bibinfo{year}{1990}).

\bibitem[{\citenamefont{Bar-Jossef et~al.}(2004)\citenamefont{Bar-Jossef,
  Jayram, and Kerenidis}}]{BJK:stoc04}
\bibinfo{author}{\bibfnamefont{Z.}~\bibnamefont{Bar-Jossef}},
  \bibinfo{author}{\bibfnamefont{T.~S.} \bibnamefont{Jayram}},
  \bibnamefont{and}
  \bibinfo{author}{\bibfnamefont{I.}~\bibnamefont{Kerenidis}}, in
  \emph{\bibinfo{booktitle}{Proceedings of the 36th Annual ACM Symposium on
  Theory of Computing}} (\bibinfo{year}{2004}), pp. \bibinfo{pages}{128--137}.

\bibitem[{\citenamefont{Buhrman et~al.}(2011)\citenamefont{Buhrman, Regev,
  Scarpa, and de~Wolf}}]{BRS:ccc11}
\bibinfo{author}{\bibfnamefont{H.}~\bibnamefont{Buhrman}},
  \bibinfo{author}{\bibfnamefont{O.}~\bibnamefont{Regev}},
  \bibinfo{author}{\bibfnamefont{G.}~\bibnamefont{Scarpa}}, \bibnamefont{and}
  \bibinfo{author}{\bibfnamefont{R.}~\bibnamefont{de~Wolf}}, in
  \emph{\bibinfo{booktitle}{Proceedings of the 26th IEEE Annual Conference on
  Computational Complexity}} (\bibinfo{year}{2011}), pp.
  \bibinfo{pages}{157--166}.

\bibitem[{\citenamefont{Aumann}(1974)}]{aumann:jme74}
\bibinfo{author}{\bibfnamefont{R.}~\bibnamefont{Aumann}},
  \bibinfo{journal}{Journal of Mathematical Economics}
  \textbf{\bibinfo{volume}{1}}, \bibinfo{pages}{67} (\bibinfo{year}{1974}).

\bibitem[{\citenamefont{Lo and Chau}(1998)}]{lochau98}
\bibinfo{author}{\bibfnamefont{H.-K.} \bibnamefont{Lo}} \bibnamefont{and}
  \bibinfo{author}{\bibfnamefont{H.~F.} \bibnamefont{Chau}},
  \bibinfo{journal}{\emph{Physica D}} \textbf{\bibinfo{volume}{120}},
  \bibinfo{pages}{177} (\bibinfo{year}{1998}).

\bibitem[{\citenamefont{Mayers}(1997)}]{mayers}
\bibinfo{author}{\bibfnamefont{D.}~\bibnamefont{Mayers}},
  \bibinfo{journal}{\emph{Phys. Rev. Lett.}} \textbf{\bibinfo{volume}{78(17)}},
  \bibinfo{pages}{3414} (\bibinfo{year}{1997}).

\bibitem[{\citenamefont{Cleve et~al.}(2004)\citenamefont{Cleve, Hoyer, Toner,
  and Watrous}}]{Cleve:CC04}
\bibinfo{author}{\bibfnamefont{R.}~\bibnamefont{Cleve}},
  \bibinfo{author}{\bibfnamefont{P.}~\bibnamefont{Hoyer}},
  \bibinfo{author}{\bibfnamefont{B.}~\bibnamefont{Toner}}, \bibnamefont{and}
  \bibinfo{author}{\bibfnamefont{J.}~\bibnamefont{Watrous}}, in
  \emph{\bibinfo{booktitle}{Proceedings of the 19th IEEE Annual Conference on
  Computational Complexity}} (\bibinfo{year}{2004}), pp.
  \bibinfo{pages}{236--249}.

\bibitem[{\citenamefont{Navascues et~al.}(2007)\citenamefont{Navascues,
  Pironio, and Acin}}]{NPA:prl2007}
\bibinfo{author}{\bibfnamefont{M.}~\bibnamefont{Navascues}},
  \bibinfo{author}{\bibfnamefont{S.}~\bibnamefont{Pironio}}, \bibnamefont{and}
  \bibinfo{author}{\bibfnamefont{A.}~\bibnamefont{Acin}},
  \bibinfo{journal}{Phys. Rev. Lett.} \textbf{\bibinfo{volume}{98}},
  \bibinfo{pages}{010401} (\bibinfo{year}{2007}).

\bibitem[{\citenamefont{Navascues et~al.}(2008)\citenamefont{Navascues,
  Pironio, and Acin}}]{NPA:njp2008}
\bibinfo{author}{\bibfnamefont{M.}~\bibnamefont{Navascues}},
  \bibinfo{author}{\bibfnamefont{S.}~\bibnamefont{Pironio}}, \bibnamefont{and}
  \bibinfo{author}{\bibfnamefont{A.}~\bibnamefont{Acin}}, \bibinfo{journal}{New
  J. Phys.} \textbf{\bibinfo{volume}{10}}, \bibinfo{pages}{073013}
  (\bibinfo{year}{2008}).

\bibitem[{\citenamefont{Pironio et~al.}(2010)\citenamefont{Pironio, Navascues,
  and Acin}}]{PNA:siam2010}
\bibinfo{author}{\bibfnamefont{S.}~\bibnamefont{Pironio}},
  \bibinfo{author}{\bibfnamefont{M.}~\bibnamefont{Navascues}},
  \bibnamefont{and} \bibinfo{author}{\bibfnamefont{A.}~\bibnamefont{Acin}},
  \bibinfo{journal}{SIAM Journal on Optimization}
  \textbf{\bibinfo{volume}{20}}, \bibinfo{pages}{2157} (\bibinfo{year}{2010}).

\bibitem[{\citenamefont{Kempe et~al.}(2010)\citenamefont{Kempe, Regev, and
  Toner}}]{KRT:siam2010}
\bibinfo{author}{\bibfnamefont{J.}~\bibnamefont{Kempe}},
  \bibinfo{author}{\bibfnamefont{O.}~\bibnamefont{Regev}}, \bibnamefont{and}
  \bibinfo{author}{\bibfnamefont{B.}~\bibnamefont{Toner}},
  \bibinfo{journal}{SIAM Journal on Computing} \textbf{\bibinfo{volume}{39}},
  \bibinfo{pages}{3207} (\bibinfo{year}{2010}).

\bibitem[{\citenamefont{http://www.qutools.com/products/quED}()}]{QuTools}
\bibinfo{author}{\bibnamefont{http://www.qutools.com/products/quED}}.

\bibitem[{\citenamefont{Benjamin and Hayden}(2001)}]{benhay:pra01}
\bibinfo{author}{\bibfnamefont{S.~C.} \bibnamefont{Benjamin}} \bibnamefont{and}
  \bibinfo{author}{\bibfnamefont{P.~M.} \bibnamefont{Hayden}},
  \bibinfo{journal}{Phys. Rev. A} \textbf{\bibinfo{volume}{64}},
  \bibinfo{pages}{030301} (\bibinfo{year}{2001}).

\end{thebibliography}
\end{document}